\documentclass[twocolumn,showpacs,amsmath,amssymb]{revtex4}

\usepackage{graphicx}
\usepackage{dcolumn}
\usepackage{bm}
\usepackage{times,mathptm}
\newcommand{\be}{\begin{equation}}
\newcommand{\ee}{\end{equation}}
\newcommand{\bea}{\begin{eqnarray}}
\newcommand{\eea}{\end{eqnarray}}

\newcommand{\RR}{\rangle}
\newcommand{\LL}{\langle}

\begin{document}


\title{Scalable register initialization for quantum computing in an optical lattice\\}

\author{Gavin K. Brennen$^{1,2}$, Guido Pupillo$^{1,2}$, Ana Maria Rey$^{1,2}$, Charles W. Clark$^1$, and  Carl J. Williams$^1$}
\affiliation{$^1$
National Institute of Standards and Technology,
Gaithersburg, Maryland 20899 \\$^2$Department of Physics, University of Maryland, College Park, Maryland 20742}

\date{\today}
\begin{abstract}
The Mott insulator state created by loading an atomic Bose-Einstein condensate (BEC) into an optical lattice may be used as a means to prepare a register of atomic qubits in a quantum computer.  Such architecture requires a lattice commensurately filled with atoms, which corresponds to the insulator state only in the limit of zero inter-well tunneling.  We show that a lattice with spatial inhomogeneity created by a quadratic magnetic trapping potential can be used to isolate a subspace in the center which is impervious to hole-hoping.  Components of the wavefunction with more than one atom in any well can be projected out by selective measurement on a molecular photo-associative transition.  Maintaining the molecular coupling induces a quantum Zeno effect that can sustain a commensurately filled register for the duration of a quantum computation. 
\end{abstract}


\maketitle

In the past decade, tremendous progress has been made in the field of atomic physics toward the creation of macroscopic quantum states such as Bose-Einstein condensates on one hand \cite{Ketterle} and precise control of single and few coupled atoms on the other, for example in optical lattices \cite{Denschlag}.  Several years ago it was proposed \cite{Jaksch} to marry these advances by loading an optical lattice from an atomic BEC.  If one begins with a superfluid-like BEC and adiabatically turns on an optical lattice potential, the system will undego a phase transition to the Mott insulator (MI) state (characterized by the same number of atoms in each lattice well) when the intra-well interaction energy is much greater than the inter-well tunneling.\cite{Fisher}.  Recent experiments \cite{Greiner} have demonstrated the Mott insulator phase transition in a magnetically confined three dimensional optical lattice, with an average filling factor of two atoms per well.  In this paper we will show how the use of an quadratic trapping potential and selective measurement of atomic pairs allows for the MI transition to become a robust mechanism for quantum register initialization.

It was recognized early on that the MI transition might be an efficient way to initialize a register of atomic qubits in an optical lattice for use in quantum information processing.  A key advantage of loading from a BEC is the availability of an initially high phase space density which can be frozen to the MI state with atoms occupying every lattice site.  When the lattice is loaded such that only the lowest vibrational state of each lattice well is occupied, the system is well described by the Bose-Hubbard Hamiltonian:  
\be
H_{BH}=\sum_{j}\epsilon(j)n_{j}-J(a_{j}^{\dagger}a_{j+1}+a_{j+1}^{\dagger}a_{j})+\frac{U}{2} n_j(n_j-1)\;
\ee
Here $a_{j}$ are the bosonic annihilation operators and $n_{j}=a_j^{\dagger}a_j$ the number operators for an atom in the lowest vibrational state of lattice well $j$.  The energy offset at each lattice site is $\epsilon(j)$ which models a continuously varying external potential.  The energies $J$ and $U$ are the tunneling and on-site interaction energies respectively.  In the tight-binding model, the nearest neighbor tunneling energy $J$ is defined as one fourth the band width of the lowest occupied band.  For tunneling through a potential barrier given by $V(x)=V\cos^{2}(kx)$, the tunneling rate is closely approximated by \cite{MF} $J/\hbar=4/(\sqrt{\pi}\hbar)E_{R}(V/E_{R})^{3/4}e^{-2\sqrt{V/E_{R}}}$,
where the recoil energy is $E_{R}=(\hbar k)^2/2m$ (m$=$atomic mass).  The on site interaction is a result of the ground state collisions described by the s-wave scattering length $a_{s}$ between two atoms each in the motional state $\phi({\bf x})$ and is given by $U=\frac{4\pi a_{s}\hbar^2}{m}\int d{\bf x}|\phi({\bf x})|^4$.
 
For the homogeneous system $(\epsilon(j)=0\ \forall j)$ of fixed extent, the behavior of the system is uniquely described by the ratio $U/J$ which for trapping in an optical lattice decreases exponentially with the trap depth $V$.  While our results are applicable to higher dimensions, henceforth we assume a three dimensional lattice with tight transverse confinement and tunneling dynamics along one dimension only.  For the homogeneous system, only commensurate fillings give rise to a MI transition.  However, one should note that an adiabatic transfer mechanism between two sublevels of each atom  may be used to fix nonuniform filling \cite{Zoller}.  A caveat is that even with unit filling, the MI state still carries small but non-zero number fluctuations which provide a small residual coherence across the system that scales as the number of trapped atoms \cite{Guido,Roberts}.  Applying first order perturbation theory in $H_{BH}$, the ground state for $N$ atoms in $M$ wells in one dimension with $N=M$ is approximately
\be
|\Psi_{g}\RR=\alpha(|T\RR+2\sqrt{N}J/U|S\RR),\;
\ee
where the normalization constant is $\alpha=(1+4N(J/U)^2)^{-1/2}$.  Here the unit filled target state is $|T\RR=\prod_{j=1}^{N}a_{j}^{\dagger}|0\RR$ and the symmetrized state, assuming periodic boundary conditions $(j+M\equiv j)$, is $|S\RR=1/\sqrt{2N}\sum_{j=1}^{2N}(a_{j+1}^{\dagger}a_{j}+a_{j}^{\dagger}a_{j+1})|T\RR$.  The energy of the ground state is approximately $E_g=-4NJ^2/U$.  In general, a homogeneous lattice will have a probability of approximately $1/M$ of being commensurately filled and thus does not provide a robust mechanism for initializing a quantum computer.

We therefore propose to use an inhomogeneous lattice with open boundaries created by a weak quadratic magnetic trap that acts to collect atoms near the center of the trap and leaves empty wells (holes) at the edges.  For our analysis we assume a one dimensional optical lattice, with $N<M$ \cite{footnoteII}, in the presence of a weak magnetic trap with oscillation frequency $\omega_T$.  The characteristic trap energy scale $\delta=m/2(\pi/k)^2 \omega_T^2$ is defined so that $\epsilon(j)=\delta j^2$.  We stipulate that the on site interaction energy be larger than the trapping energy of the most externally trapped atom,  or $U>\epsilon((N-1)/2)$, in order to inhibit multiple atom occupation in any well.  The register is defined by a physical subspace ${\mathcal R}$ comprising a number of wells $n<N$ in the center region of the trap.  The barrier space flanking ${\mathcal R}$ will act to suppress  percolation of holes from the edges to the center.  The estimated probability for holes in ${\mathcal R}$ due to tunneling through the barrier is $p_h\approx\prod_{j=(n-1)/2}^{(N-1)/2}(J/\delta(2j+1))^2=(J/2\delta)^{N-n+2}(\Gamma[n/2]/\Gamma[N/2+1])^2$, which is negligible provided the barrier region is sufficiently large and $J/n\delta<1$.

When expanded in the Fock state basis, the ground state of the register has amplitude in those states with holes neighboring atomic pairs, analogous to the homogeneous case.  We describe a protocol which projects out these components by a null result from selective measurement of atomic pairs within any lattice site.  This measurement detects population on an excited molecular state and can be made with high efficiency.  Once the unit filled state is reached with high confidence, continuing the measurement will maintain this state by virtue of the quantum Zeno effect.  

The measurement will map the register from the ground state $|\Psi_g\RR$ to the unit filled target state which is not an eigenstate of $H_{BH}$.  To describe the dynamics in the register during the measurement, it is convenient to use the following incomplete basis over ${\mathcal R}$:
\be
|T\RR=\prod_{j=-(n-1)/2}^{(n-1)/2}a_{j}^{\dagger}|0\RR,|S_{j}^+\RR=\frac{a_{j}^{\dagger}a_{j+1}}{\sqrt{2}}|T\RR,|S_{j}^-\RR=\frac{a_{j+1}^{\dagger}a_{j}}{\sqrt{2}}|T\RR.\;
\label{basis}
\ee
For each $j$ the states $|S_j^{\pm}\RR$ are distinguished by the two energetically distinct orientations of an atomic pair and its neighboring hole with energies, $E(S_j^{\pm})=U(1\mp \frac{\delta}{U} (2 j-1))$.  The target state $|T\RR$ defines the zero of energy.  The $|T\RR$ and the $|S_j^{\pm}\RR$ states are coupled to first order in $H_{BH}$ and they span the reduced state in ${\mathcal R}$ of the ground state of the total system.  

In the limit of large $n$, the dynamics of the register is restricted to the basis of Eq. \ref{basis}.  This argument is understood by comparison to the dynamics in the homogeneous system.  In the latter, the state with the largest coupling from the target state is the symmetrized $|S\RR$ with coupling matrix element $\LL T|H_{BH}|S\RR=-2\sqrt{n}J$.  The state $|S\RR$ itself couples to a symmetrized state $|S^{\prime}\RR$ of all Fock states with a one site separation between the atomic pair and the hole:  $|S^{\prime}\RR=1/\sqrt{2n}\sum_{j}(a_{j}^{\dagger}a_{j+2}+a_{j+2}^{\dagger}a_{j})|T\RR$.  The coupling between these states is $\LL S^{\prime}|H_{BH}|S\RR=-3J$.  The dynamics on time scales $t<1/J$ are therefore constrained to the subspace $\{|T\RR,|S\RR\}$ when $n\gg1$.  In the inhomogeneous case, the degeneracy is absent between states with neighboring pairs and holes, $\{|S_j^{\pm}\RR\}$ and states where pairs and holes are separated.  In this off resonant situation, the coupling to states outside the restricted subspace can only be smaller than in the homogeneous case.

\begin{figure}
\begin{center}
\includegraphics[scale=0.4]{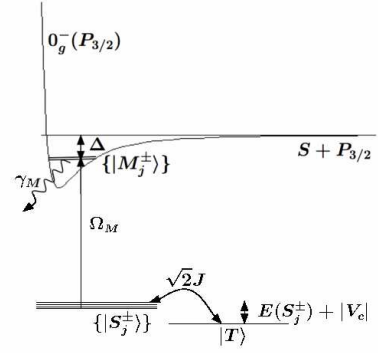}
\caption{\label{fig:1}Schematic of the relevant couplings in the problem.  The unit filled state $|T\RR$ describing a target quantum register and the states $|S_j^{\pm}\RR$ having one doubly occupied lattice site and a neighboring hole are coupled to first order in $H_{BH}$.  A catalysis laser couples states $|S_j^{\pm}\RR$  to the excited states $|M_j^{\pm}\RR$ having a bound molecule at the doubly occupied site.  The bound states quickly decay and give the possibility of monitoring population in the ``faulty" register states $|S_j^{\pm}\RR$.}
\end{center}
\end{figure}

For the measurement, we choose a catalysis laser that is on resonance from the ground state of two atoms in a single well to a bound state $\nu$ of a dipole-dipole coupled molecular $S+P$ state.  The bound state is chosen such that the catalysis laser is far off resonance from other bound states and repulsive potentials, see Fig.\ref{fig:1}.  For our many body system, we adopt the set of many body states $\{|M_j^{\pm}\RR=(1/\sqrt{2})b_{j+(1\mp1)/2}^{\dagger}a_{j+(1\mp1)/2}^2|S_j^{\pm}\RR\}$, where $b^{\dagger}$ is the creation operator for a molecule in the bound state $\nu$.  These states describe $n-2$ atoms trapped in the lattice and a single molecule at site $j+(1\mp1)/2$, with dipole-dipole coupling energy $\LL M_j^{\pm}|H_{dd}|M_j^{\pm}\RR=\hbar\omega_{\nu}$.  The free atomic Hamiltonian for $n$ atoms is $H_{A}=\sum_{j}\hbar \omega_{eg}|e_j\RR\LL e_j|$ where $|e_j\RR(|g_j\RR)$ denotes the excited(ground) state for an atom at site $j$.  Hereafter, we work in units with the numerical value of $\hbar$ equal to $1$ meaning energies are understood as being in units of inverse time.  The ``bare" energy Hamiltonian $H_{0}$ including coupling in the restricted basis of $H_{BH}$ is then:
\be
\begin{array}{lll}    
H_{0}&=&H_A+H_{dd}+H_{BH}\\
&=&\omega_{eg}\sum_{j}|e_j\RR\LL e_j|+\sum_{j,\pm}E(S_{j}^{\pm})|S_{j}^{\pm}\rangle\langle S_{j}^{\pm}|\\
& &+(\omega_{\nu}+E(S_j^{\pm})-U)|M_j^{\pm}\RR\LL M_j^{\pm}|\\
& &-\sqrt{2}J\sum_{j,\pm}(|S_j^{\pm}\RR\LL T|+|T\RR\LL S_j^{\pm}|),\;
\end{array}
\ee

Under the atom laser interaction, $H_{AL}$, the ground and excited state of each atom is coupled as is each many body state $|M_{j}^{\pm}\RR$ and its corresponding ground state $|S_j^{\pm}\RR$.  In the rotating wave approximation, the interaction is: 
\be
\begin{array}{lll}
H_{AL}&=&\frac{\Omega_A}{2}\sum_j (e^{-i\omega_L t}|e_j\RR\LL g_j|+h.c.)\\
& &+\frac{\Omega_M}{2}\sum_{j,\pm} (e^{-i\omega_L t}|M_j^{\pm}\RR\LL S_j^{\pm}|+h.c.),\;
\end{array}
\ee
where $\Omega_{A(M)}$ are the atomic (molecular) Rabi frequencies, related by $\Omega_M=\sqrt{F_{\nu}}\Omega_A$, where $F_{\nu}$ is the Franck-Condon factor equaling the spatial overlap between the relative co\"ordinate wavefunction describing two ground electronic state atoms trapped in a single lattice well and the molecular bound state $\nu$.  For bound states of interest, such as the long range bound states of the $0_g^-(P_{3/2})$ potential \cite{Fioretti}, the catalysis detuning from {\it atomic} resonance, $\Delta=\omega_L-\omega_{eg}$, is several thousands linewidths meaning the atomic saturation is low $s_A=(\Omega_A^2/2\Delta^2)\ll1$.  In this case, the excited atomic states can be eliminated and the total Hamiltonian $H_0+H_{AL}$ in the rotating frame is,
\be
\begin{array}{lll}
H_I&=&\sum_{j,\pm}((|V_c|+E(S_j^{\pm}))|S_j^{\pm}\RR\LL S_j^{\pm}|)\\
& &+(|V_{c}|+E(S_j^{\pm})-U)|M_j^{\pm}\RR\LL M_j^{\pm}|\\
& &-\sqrt{2}J(|S_j^{\pm}\RR\LL T|+|T\RR\LL S_j^{\pm}|)\\
& &+\frac{\Omega_M}{2}(|M_j^{\pm}\RR\LL S_j^{\pm}|+|S_j^{\pm}\RR\LL M_j^{\pm}|)).\;
\end{array}
\ee
Here, the differential light shift seen by the $|T\RR$ state and the $|S_j^{\pm}\RR$ and $|M_j^{\pm}\RR$ states is equal to twice the single atom light shift:  $V_{c}=\Delta s_{A}$.  
Any population in the bound molecular states will decay at a rate $\gamma_M\approx 2 \Gamma$, where $\Gamma$ is the single atom decay rate.  Because the excited state linewidth is greater than the lattice depth, the decay products will typically not be trapped and the system evolves according to a trace non-preserving master equation:
\be
\dot{\rho}=-i[H_I,\rho]-\gamma_M/2\sum_{j,\pm} (|M_j^{\pm}\RR\LL M_j^{\pm}|\rho+\rho |M_j^{\pm}\RR\LL M_j^{\pm}|).\;
\label{master}
\ee
We have ignored spontaneous emission due to decay from the single atom excited states at a rate $s_A \Gamma\ll\Gamma$ per atom.  
For time scales $1/\gamma_{M}\ll t\ll1/(U+|V_c|)$, the excited state coherences can be solved for.  To second order in $U/\gamma_{M},|V_c|/\gamma_M$ and first order in $J/\gamma_M$ they are
\be
\begin{array}{lll}
\rho_{M_{j}^{\pm},T}(t)&=&-i \frac{\Omega_{M}\gamma_{M}/4}{(|V_c|+E(S_j^{\pm})-U)^2+(\gamma_{M}/2)^2}\rho_{S_{j}^{\pm},T}(t)\\
\rho_{M_{j}^{\pm},M_{j}^{\pm}}(t)&=&\frac{\Omega_{M}^2/4}{U^2+(\gamma_{M}/2)^2}\rho_{S_{j}^{\pm},S_{j}^{\pm}}(t)\\
\rho_{M_{j}^{\pm},S_{j}^{\pm}}(t)&=&-i\frac{\Omega_{M} \gamma/4}{U^2+(\gamma_{M}/2)^2}\rho_{S_{j}^{\pm},S_{j}^{\pm}}(t).\;
\end{array}
\ee
Inserting these expressions back into the equations for the dynamics in the ground state we have
\be
\begin{array}{lll}
\dot{\rho}_{S_{j}^{\pm},T}&=&-i(E(S_{j}^{\pm})+|V_{c}|)\rho_{S_{j}^{\pm},T}+i\sqrt{2} J(\rho_{T,T}-\rho_{S_{j}^{\pm},S_{j}^{\pm}})\\
& &-\frac{\Omega_{M}^2\gamma_{M}/8}{(|V_c|+E(S_j^{\pm})-U)^2+(\gamma_{M}/2)^2}\rho_{S_{j}^{\pm},T}\\
\dot{\rho}_{T,T}&=&i\sqrt{2}J(\rho_{S_{j}^{\pm},T}-\rho_{T,S_{j}^{\pm}})\\
\dot{\rho}_{S_{j}^{\pm},S_{j}^{\pm}}&=&-i\sqrt{2}J(\rho_{S_{j}^{\pm},T}-\rho_{T,S_{j}^{\pm}})-\frac{\Omega_{M}^2 \gamma_M/4}{U^2+(\gamma_{M}/2)^2} \rho_{S_{j}^{\pm},S_{j}^{\pm}}.\;
\end{array}
\label{deneqs}
\ee
These equations describe the Bose-Hubbard coupled states with a decay in population of each state with an atomic pair at a rate $2\kappa=\Omega_M^2\gamma_M/(4 (U^2+(\gamma_M/2)^2)$, and a decay of the coherences between each of these states and the $|T\RR$ state at approximately the rate $\kappa$.  

This type of evolution characterized by measurement induced phase damping was studied extensively by Gagen and Milburn \cite{Milburn}.  We now show that our system can satisfy the conditions for this effect and in particular can be driven to the $|T\RR$ state by monitoring the environment for a signature of decay from the molecular bound state.

For the inhomogeneous system, the state $|T\RR$ couples to $2n$ distinguishable states $|S_j^{\pm}\RR$.  However, we can define an effective Rabi frequency between the subspace $|T\RR$ and the subspace spanned by $\{|S_J^{\pm}\RR\}$.  This frequency is close to the coupling matrix element between the state $|T\RR$ and the state $|S\RR$ in the homogeneous system, namely $2\sqrt{n}J$.  The coherences between the two subspaces decay at a rate $\kappa$, and the population in the subspace $\{|S_j^{\pm}\RR\}$ decays at a rate $2\kappa$.  The ``good" measurement regime as derived in \cite{Milburn} is then:
\be
\frac{\Omega_M}{\gamma_M}\ll 1 < \frac{\kappa}{2 \sqrt{n} J}.\;
\ee
The left side inequality ensures that the excited states $|M_j^{\pm}\RR$ are weakly populated (equivalent to the condition for adiabatic elimination of these states).  The right side inequality ensures that measurement is sufficiently strong to damp coherences on the time scale that they develop due to tunneling.

The limiting quantity that determines the decay rate of the weakly saturated molecular states and hence the measurement strength is the Franck-Condon factor $F_{\nu}$.  It is calculated for bound-bound transitions using the reflection approximation of Julienne \cite{Julienne} where we solve for the ground state relative co\"ordinate wavefunction for two atoms in a lattice well using a pseudo potential appropriate for $^{87}Rb$.  We choose to couple to the $\nu=17$ bound state of the $0_g^-(P_{3/2})$ potential which is at an energy $\Delta=-6.85\times10^4 \Gamma$ from dissociation.  For a lattice with wavelength  $785\ \mbox{nm}$ and transverse and parallel confinements $V^{\perp}=38.5 E_R$, $V^{||}=22 E_R$, the result is $F_{\nu}\approx 5 \times 10^{-7}$.  Given this confinement, the on-site interaction using $a_s=5.6\ \mbox{nm}$ is $U=3.574\mbox{kHz}$.  Choosing an experimentally reasonable atomic Rabi frequency $\Omega_A=25\Gamma$, where $\Gamma=2\pi\times6.065\mbox{MHz}$, we find $\kappa\approx 0.13 U$.  Here the atomic scattering due to the catalysis laser is $s_A \Gamma\approx 6.7\times10^{-8} \Gamma$ per atom and the off resonant light shift is $|V_c|=15.5 U$. 

By way of example we define a one dimensional register ${\mathcal R}$with $501$ atoms that resides inside a lattice filled with $N=551$ atoms.  An external magnetic trapping frequency of $\omega_T=2\pi\times 8$Hz ensures that the last occupied well has an energy $\epsilon((N-1)/2)=0.9 U$.  We note that the probability for tunneling of holes from the edges is negligible as $J/(n\delta)=0.28$.  In practice it is not important to know the exact number of atoms in the lattice as long as the trap strength is chosen such that, given the uncertainty in the number of atoms, the constraint $\epsilon((N-1)/2)< U$ is always satisfied.  These parameters fix the ratio $U/J=500$ and the measurement strength is therefore $\kappa/2\sqrt{n}J\approx 1.5$.

\begin{figure}
\begin{center}
\includegraphics[scale=0.45]{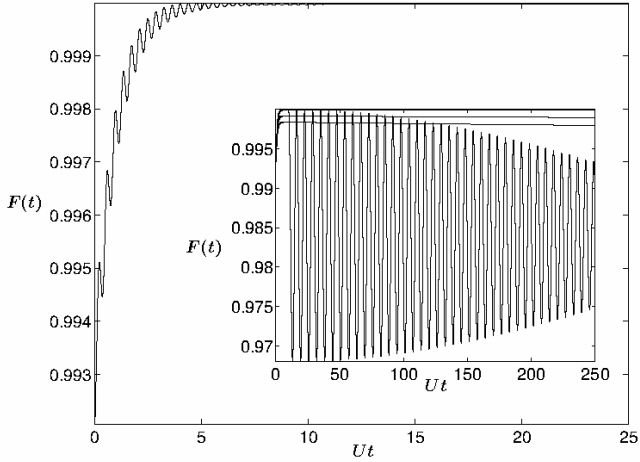}
\caption{\label{fig:2}Target state fidelity, $F(t)=\rho_{T,T}(t)$, for $U/J=500$ and $n=500$ atoms.  The main plot shows a quantum trajectory for a state that begins in the ground state of the Bose-Hubbard Hamiltonian and converges to the unit filled state $|T\RR$ by a null result on a selective measurement.  The time scale for saturating the target state for the parameters here is $t_{sat}\approx\kappa^{-1}=7.7/U$.  The inset shows the evolution for three detector efficiencies $(\eta=1.0,0.9,0.8)$ over a longer time scale.  Also shown is the oscillatory dynamics with frequency $U$ if the measurement is turned off after the target state is reached.}
\end{center}
\end{figure}

When the environment is monitored, for instance by looking for photon scattering from the bound molecular state, the evolution of ground states can be modeled using quantum trajectories. For our simulation, the ground state wavefunction $|\psi(t)\RR=c_T(t)|T\RR+\sum_{j,\pm}c_{S_j^{\pm}}(t)|S_j^{\pm}\RR$ is updated according to the non-Hermitian Hamiltonian $H=H_{I}-i\gamma_M/2\sum_{j,\pm}|M_j^{\pm}\RR\LL M_j^{\pm}|$.   A quantum trajectory corresponding to a null measurement result converges to the target state $|T\RR$ and freezes the state there as demonstrated in Fig. \ref{fig:2} (for a similar effect with ions see \cite{Beige}).   The preparation time scale is $t_{prep}=1/\kappa$.  The probability that the preparation fails, i.e. the system decays from the excited states is $p_{fail}=1-\rho_{T,T}(0)$.

Real experiments will be constrained to finite detector efficiencies $\eta$.  For $\eta=0$, corresponding to nonselective measurement, the system dynamics evolve according to Eq. \ref{deneqs}.  If we represent the dynamics of the system as a pseudo two state system $|T\RR$ and $|S\RR$, with an average energy splitting $U+|V_c|$, the equations of motion for the pseudo Bloch vector are:
\be
\begin{array}{lll}
\dot{u}&=&(U+|V_c|)v-\kappa u\\
\dot{v}&=&-\kappa v-(U+|V_c|)u-\sqrt{2n}Jw\\
\dot{w}&=&-\kappa(x+w)+4\sqrt{2n}Jv\\
\dot{x}&=&-\kappa(x+w),\;
\end{array}
\ee
where $u=\mbox{Re}[\rho_{S,T}]$, $v=\mbox{Im}[\rho_{S,T}]$, $w=\rho_{S,S}-\rho_{T,T}$, and the decreasing norm is $x=\mbox{Tr}[\rho]$.  After a period $1/|V_c|$, the coherences approach steady state, and the target state population, assuming $\rho_{T,T}(0)\approx 1$ and $\kappa/2 \sqrt{n}J >1$, is
\be
\rho^{ns}_{T,T}(t)=\rho_{T,T}(0)e^{-(8nJ^2\kappa t)/((U+|V_c|)^2+\kappa^2)}.\;
\ee
This behavior exhibits the continuous quantum Zeno effect in the limit $\kappa/2 \sqrt{n}J \rightarrow \infty$.  Even for moderately strong measurements, if the initial state is close to the target state then the deviation time is long compared to $1/U$.  For finite efficiencies, we can express the approximate fidelity to be in the target state.  Assuming a null measurement result, for times $t>t_{prep},1/|V_c|$, it is $F(\eta,t)=\rho_{T,T}(t)=\eta+(1-\eta)\rho^{ns}_{T,T}(t)$.  In practice, high detection efficiencies may be obtainable by applying a second photo ionizing laser on resonance with the molecular bound state and monitoring the emission of ions.               
    
If the catalysis field is turned off after the target state is reached, the system will freely evolve according to $H_{BH}$.  The general time dependent state in the restricted basis is:
\be
|\Psi(t)\RR=c_T(t)|T\RR+\sum_{j,\pm}e^{-iE(S_j^{\pm})t}c_{S_j^{\pm}}(t)|S_j^{\pm}\RR.\;
\ee
\noindent
Assuming the initial state $|\Psi(0)\RR=|T\RR$, the amplitudes $c_T,c_{S_j^{\pm}}$ evolve according to the Schr\"odinger equation, $i\frac{\partial}{\partial t}|\Psi(t)\RR=H_{BH}|\Psi(t)\RR$.  Using first order perturbation theory the time dependent fidelity to be in the target state $F(t)=|\LL T|\Psi(t)\RR|^2$ can be estimated under the assumption that $\delta/U\ll1$.   For times $t<1/J$ we find
\be
F(t)=1-8(J/U)^2(n-\cos(Ut)(1+\sin(\delta (n-1)t)/\sin(\delta t)).\;
\label{fidelity}
\ee
This solution compares well with exact numerical simulations for a small number of atoms $(N,M<8)$ and simulations for larger $N=M$ using a restricted basis set of dimension $N(N-1)+1$ consisting of states with at most one site with two atoms.  Note that the time averaged deviation from perfect fidelity $1-\langle F(t)\rangle=8n(J/U)^2$ is twice as bad as the deviation if the system were left in the ground state $|\Psi_g\RR$.  Because the Bose-Hubbard Hamiltonian is intrinsic to trapped bosons in an optical lattice, other {\it dispersive} techniques for initializing a register in a lattice such as Raman side-band cooling \cite{Jessen}, and phase space compression \cite{Weiss}, if not corrected, will suffer from the same loss of fidelity as described by Eq. \ref{fidelity}.     


In summary, we have shown that efforts to prepare a register of atomic qubits in an optical lattice suffer from errors inherent in the underlying many body dynamics.  We have introduced a protocol that addresses this issue to make the MI transition a robust mechanism for initialization.  While we have presented the idea in the context of one dimensional dynamics, the method is also applicable to higher dimensions, which is the relevant regime for scalability.
We appreciate helpful discussions with Almut Beige, Paul Julienne, and Eite Tiesinga.  This research was supported in part by ARDA/NSA.


\begin{references}

\bibitem{Ketterle}  M.R. Andrews {\it et al.}, Science {\bf 275}, 637 (1997).

\bibitem{Denschlag} J.H. Denschlag {\it et al.}, J. Phys. B {\bf 35}, 3095 (2002).

\bibitem{Jaksch} D. Jaksch {\it et al.}, Phys. Rev. Lett. {\bf 81}, 3108 (1998).

\bibitem{Fisher} M.P.A. Fisher, {\it et al.}, Phys. Rev. B {\bf 40}, 546 (1989).

\bibitem{Greiner} M. Greiner {\it et al.}, Nature, {\bf 415}, 39 (2002).

\bibitem{MF} Mathieusche Funktionen und Sphaeroidfunktionen, J. Meixner and F.W. Schaefke (Springer Verlag, Berlin, Germany, 1954). 

\bibitem{Zoller} P. Rabl {\it et al.}, Phys. Rev. Lett. {\bf 91}, 110403 (2003).

\bibitem{Roberts} D.C. Roberts and K. Burnett, Phys. Rev. Lett. {\bf 90}, 150401 (2003).

\bibitem{Guido} G. Pupillo, E. Tiesinga, and C. J. Williams, Phys. Rev. A (accepted); cond-mat/0308062.


\bibitem{footnoteII} For convenience we choose $N$ odd and fix the site index $j=0$ at the minimum of the magnetic trap.

\bibitem{Fioretti} A. Fioretti {\it et al.}, Eur. Phys. J. D {\bf 15}, 189 (2001).

\bibitem{Milburn} M.J. Gagen and G.J. Milburn, Phys. Rev. A {\bf 47}, 1467 (1993).


\bibitem{Beige} A. Beige {\it et al.}, Phys. Rev. Lett. {\bf 85}, 1762 (2000).

\bibitem{Julienne} P.S. Julienne, J. Res. Natl. Inst. Stand. Technol. {\bf 101}, 487 (1996).

\bibitem{Jessen} S.E. Hamann {\it et al.}, Phys. Rev. Lett. {\bf 80}, 4149 (1998).

\bibitem{Weiss} D.J. Han, M.T. DePue, and D.S. Weiss, Phys. Rev. A {\bf 63}, 023405 (2001).




\end{references}

\end{document}